# Climate Trends of Tropical Cyclone Intensity and Energy Extremes Revealed by Deep Learning

Buo-Fu Chen,

**Center for Weather Climate and Disaster Research, National Taiwan University, Taiwan**

Boyo Chen

**The University of Tokyo, Tokyo, Japan**

**and Center for Weather Climate and Disaster Research, National Taiwan University, Taiwan**

Chun-Min Hsiao,

**Weather Forecast Center, Central Weather Bureau, Taiwan**

Hsu-Feng Teng

**Department of Atmospheric Sciences, National Taiwan University, Taiwan**

Cheng-Shang Lee and Hung-Chi Kuo

**Center for Weather Climate and Disaster Research, National Taiwan University, Taiwan**

**Department of Atmospheric Sciences, National Taiwan University, Taiwan**

## Contributions

B.F.C., C.S.L., and H.C.K. merged the research; C.M.H. and B.F.C. prepared the data and evaluated model performance; B.C. and B.F.C. designed the model and conducted experiments; H.F.T. and B.F.C. analyzed the TC climatology and produced the figures; all authors wrote the manuscript.

## Corresponding author

Buo-Fu Chen [e-mail: bfchen@ntu.edu.tw];

National Taiwan University, No. 1, Sec. 4, Roosevelt Road, Taipei, 10617 Taiwan

# Abstract

Anthropogenic influences have been linked to tropical cyclone (TC) poleward migration, TC extreme precipitation, and an increased proportion of major hurricanes [1, 2, 3, 4]. Understanding past TC trends and variability is critical for projecting future TC impacts on human society considering the changing climate [5]. However, past trends of TC structure/energy remain uncertain due to limited observations; subjective-analyzed and spatiotemporal-heterogeneous "best-track" datasets lead to reduced confidence in the assessed TC repose to climate change [6, 7]. Here, we use deep learning to reconstruct past "observations" and yield an objective global TC wind profile dataset during 1981–2020, facilitating a comprehensive examination of TC structure/energy. By training with uniquely labeled data integrating best tracks and numerical model analysis of 2004–2018 TCs, our model converts multichannel satellite imagery to a 0-750-km wind profile of axisymmetric surface winds. The model performance is verified to be sufficient for climate studies by comparing it to independent satellite-radar surface winds. Based on the new homogenized dataset, the major TC proportion has increased by ~13% in the past four decades. Moreover, the proportion of extremely high-energy TCs has increased by ~25%, along with an increasing trend (> one standard deviation of the 40-y variability) of the mean total energy of high-energy TCs. Although the warming ocean favors TC intensification, the TC track migration to higher latitudes and altered environments further affect TC structure/energy. This new deep learning method/dataset reveals novel trends regarding TC structure extremes and may help verify simulations/studies regarding TCs in the changing climate.

# Main

In 2018, the World Meteorological Organization announced a > 1.5°C warming of global mean sea-surface temperature (SST) since the 1980s [8, 9], with tropical SST exhibiting a robust trend of >0.5°C per decade. We face shifting climate regimes, the rapid evolution of atmospheric general circulation, expanding tropics, and increasing risk of exposure to weather extremes in this warming climate [8].

The activity of tropical cyclones (TCs), the most powerful synoptic-scale severe weather system with large social impacts [10, 11], is closely linked to ocean enthalpy and other environmental conditions, such as vertical wind shear [12, 13]. Given the warming SST, it is important to examine the anthropogenic influence on TCs through observations during the past decades [3, 4, 14]. Observational analyses of TC climate trends can inform our confidence in future TC projections under the changing climate [5].

The literature has suggested that global warming contributes to a poleward migration of TC tracks [1, 2, 3] and a slowing of forward motion [15, 16], while the TC number and TC lifetime remain approximately the same [3]. Notably, many studies have examined whether the warming SST fuels more powerful TCs [4, 14] based on the TC maximum potential intensity (MPI) theory [17, 18], in which a TC is considered a Carnot engine. They suggested that TCs may become more intense in terms of an increased fractional proportion of TCs of major hurricane severity [3, 4, 5, 14].

However, subjectively analyzed and spatiotemporally heterogeneous historical datasets, known as the best track, lead to uncertainties and reduced confidence in the assessed responses [6, 7, 19, 20]. Moreover, best tracks cannot provide sufficient data for examining the climate trends for other TC parameters in addition to intensity. Thus, neither observed anthropogenic influences on TC wind structure or overall energy in the past nor a concurred future projection have been reported [5, 21]. To this end, the lack of literature discussing TC structure in the changing climate highlights the need for comprehensive and homogenized TC structure data.

Meteorologists have a long history of observing TC intensity in terms of either minimum sea-level pressure or maximum sustained wind ($V_{max}$) [22]. Intensity is a straightforward and relatively easy-to-estimate parameter describing the local maximum of TC winds, hence setting the coordinate of studying TC climatology, development, intensification, and forecasting (e.g., the Saffir-Simpson Hurricane Wind Scale [23, 24], theory of MPI [17, 18] and TC rapid intensification [25, 26]). However, the intensity is not the best structural parameter representing TC destructive potential, TC-induced storm surge, overall energy, and feedback to larger-scale regional/climate systems [27, 28, 29, 30, 31], because knowing the TC intensity is not sufficient to determine the structure of a TC.

Numerous studies have defined and examined other structural parameters to provide a comprehensive picture of the understanding of TC wind evolution. Holland and Merill [32] and subsequent studies [33, 34] examined TC size and strength. The changes in strength and intensity seem to be independent events and are separately controlled by various internal dynamic processes [35, 31] and external forcings from the environment [36, 37, 38, 39, 40]. Furthermore, most previous studies on TC structure were conducted in the western North Pacific or Atlantic and were limited by unstable observation samples over less than ten years [41, 37, 38, 42]. There has not yet been a set of TC structure data covering the globe and exhibiting consistent quality across different basins after the satellite era; thus, global climate trends of TC structure in the past are rarely studied.

Here, we leverage the power of deep learning to estimate TC structure and calculate overall TC energy in one fell swoop. We propose a deep learning structure analysis for typhoon (DSAT) model to construct novel reanalysis datasets for TC structure research. The DSAT takes satellite imagery as input and targets the uniquely created azimuthally mean surface wind profiles within $0-750$ km of the TC; it simultaneously provides multiple TC structure parameters , including intensity, size, strength, angular momentum [38], and integrated kinetic energy (IKE) [28, 31, 27] according to the output profile. Moreover, we use high-quality but scarce polar-orbiting satellite observations to validate the accuracy of DSAT estimates. We anticipate the systematic and comprehensive DSAT "quasiobservational" dataset to establish a pivot point to reformulate the coordinate system for TC development research, which is now constructed mainly based on intensity. Then, we demonstrate the usefulness of the DSAT model in reconstructing homogenized structure data for global TCs since 1981 and first examine the climate trends of TC structure extremes.

## The Deep Learning DSAT Model

Recent studies successfully use deep learning to process TC satellite images for identifying TC formation and center positioning [43], intensity [44, 45, 46, 47] and size estimation [48, 49], and intensity forecasting [50]. These deep learning models provide more stable and accurate guidance for TC forecasting; however, they were trained against best tracks and, thus, cannot generate detailed wind profile/structure information.

This study applies deep learning to TC structure analysis. TC satellite images of infrared (IR), water vapor (WV), visible (VIS), and passive microwave rain rates (PMW) are used for training (*methods: input satellite imagery*) the DSAT model, which generates 1-D, axis-symmetric TC wind profiles within a 0−750-km radius with a 5-km radial resolution (Figs. 1a,b).

A unique labeling procedure is adopted, enabling the DSAT to end-to-end transform satellite imagery into entire TC wind profiles instead of only certain parameters [49]. Labeled data are generated by using a parametric model of Morris and Ruff [51] to integrate the intensity ($V_{max\_BT}$), radius of 34-kt winds ($R_{34\_BT}$), and radius of maximum winds ($RMW_{BT}$) from best tracks and outer TC surface winds from the ERA5 reanalysis [52] (*methods: labeled data*):

$$\vec{V}_{ERA5} = \vec{V}_{TC\_ERA5} + \vec{V}_{env\_ERA5} \tag{Eq. 1}$$

$$V_{max\_modified} = V_{max\_BT} - \overline{mag(\vec{V}_{env\_ERA5})} \tag{Eq. 2}$$

$$V_{labeled}(r) = \begin{cases} Fit(V_{max\_modified}, R_{34\_BT}, RMW_{BT}); & r < R_{34\_BT} \\ sigmoid\ linking\ formula; & R_{34\_BT} > r > R_{34\_BT} + 100 \\ \overline{mag(\vec{V}_{TC\_ERA5})}; & r > R_{34\_BT} + 100 \end{cases} \tag{Eq. 3}$$

ERA5 winds ($\vec{V}_{ERA5}$) are separated into vortex winds ($\vec{V}_{TC\_ERA5}$) and environmental flows ($\vec{V}_{env\_ERA5}$) following previous studies [53, 54]. The TC azimuthally averaged maximum wind ($V_{max\_modified}$) is obtained by subtracting the area-averaged environmental wind speed within the 500 km radius from $V_{max\_BT}$, which represents the local maxima (Eq. 2 and Fig. 1b, upper-right inset). Subsequently (Eq. 3), we use $V_{max\_modified}$, $R_{34\_BT}$ and $RMW_{BT}$ to fit the inner-core profile with the parametric model [51] while using the azimuthally mean speed of $\vec{V}_{TC\_ERA5}$ as the outer profile. The final labeled data (Fig. 1b, gold dotted line) are obtained by connecting the profile within $R_{34\_BT}$ (Fig. 1b, red line) and the outer profile (Fig. 1b, green dashed line) with a sigmoid linking formula. An exception is that for samples too weak ($V_{max}$ < 35 kt and no $R34$) to fit in the parametric model, the entire ERA5 profile ($\overline{mag(\vec{V}_{TC\_ERA5})}$) is treated as the labeled data. In summary, we collected 26917 profiles with $V_{max}$ < 35 kt and 24939 profiles of stronger $V_{max}$ for 2004−2018 global TCs to train the DSAT model (*methods: data separation*).

The DSAT model has a hybrid GAN-CNN architecture [55] (Fig. 1c), including a generative adversarial network (GAN)-[56] module and a convolutional neural network (CNN)-[57, 44] regressor. The CNN regressor takes satellite images as input, extracts essential features automatically, and then regresses these features to the labeled wind profile of azimuthally averaged winds (Fig. 1c). Furthermore, the CNN deploys polar convolutional filters [58], which makes the model focus more on the TC inner core and learn consecutive features along the tangential and radial directions, suitable for the rotative nature of TCs.

The GAN module is designed to handle temporally heterogeneous satellite data. VIS images could be critical for analyzing TC structure, but they are only available in daylight; their worse temporal coverages prohibit direct usage in training the CNN regressor. Therefore, the DSAT GAN module is trained to generate fake VIS images (VIS* in Fig. 1c) based on the all-time available IR and WV images, enabling an indirect usage of the original VIS images to train the model. Notably, in the prediction phase, the GAN module takes IR and WV to generate VIS*, and then the CNN regressor takes IR and VIS* to estimate TC wind profiles. The *methods: DSAT model* provides more details for model design and optimization.

**Fig. 1 The hybrid GAN-CNN DSAT model learns features from satellite IR, WV, and VIS images to regress to uniquely labeled TC wind profiles. The labeled data are constructed from TC best-track records, numerical-model ERA5 reanalyzed winds, and a parametric wind model.**

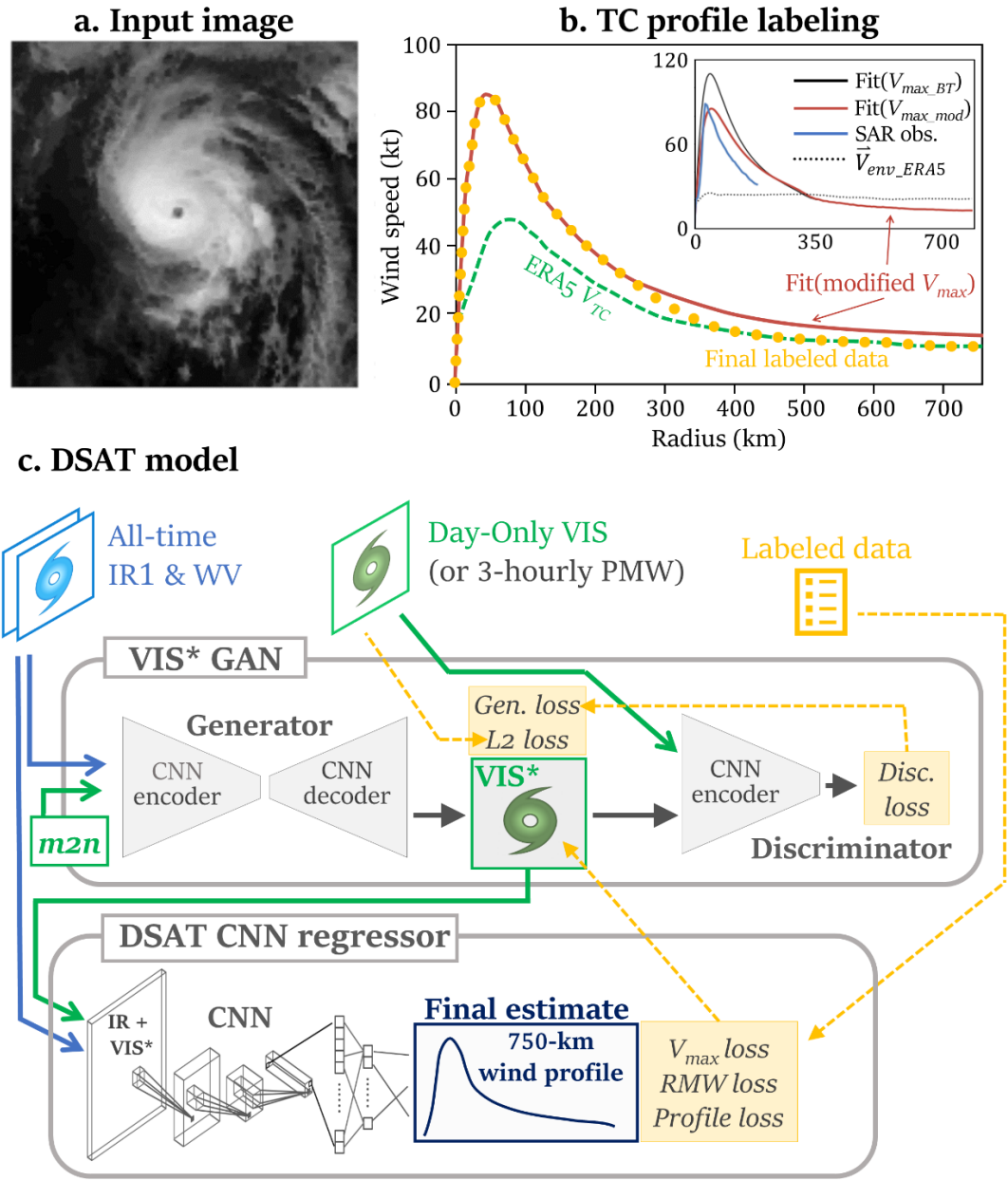

**a** An example of an input IR image. **b** The labeled data (dotted gold line) of TC 201715L at 12:00 UTC on 22 September 2017 are constructed by connecting the ERA5 profile (dashed green line) of the TC outer circulation and the TC inner profile (red solid line) within $R_{34}$ based on the parametric wind model. The parametric wind profile shown in the upper-right inset (red line) is based on the modified intensity ($V_{max\_mod}$), for which the ERA5 environmental flow (black-dotted line, $V_{env\_ERA5}$) has been removed; it exhibits comparable intensity with the SAR observation (blue line) and is better than the profile fitted based on the original best-track intensity (black line, $V_{max\_BT}$). **c** The hybrid GAN-CNN DSAT model includes a GAN module for generating mimic VIS* images based on the input IR and WV images and a DSAT CNN regressor that uses IR and VIS* images to estimate TC wind profiles within 750 km from the TC center.

# Verification

Special-collected independent verification datasets of polar-orbiting satellite observations, including Advanced Scatterometer (ASCAT) and Synthetic Aperture Radar (SAR) winds, are used to evaluate the DSAT performance (*methods: "true ground truth"*). The example of Typhoon Jondari (Fig. 2) showcases the capability of generating accurate wind profiles. The DSAT intensity and size estimates follow the ground station and satellite observations more closely than the best-track and labeled data (Figs. 2a,b). In addition, the DSAT profiles (Figs. 2c−e) follow the SAR in the inner core and the ASCAT in the outer region. Although the substantial discrepancy between the best track and the observation implies that certain errors are expected in the labeled data for some cases, the deep learning model trained on enormous amounts of data however provides opportunities to fix this issue and generate more accurate estimates.

According to statistical evaluation for 2017−2018 TCs (Extended Figs. E3c–f, E5), the DSAT model has a $V_{max}$ MAE of 12.1 (7.9) kt against SAR (labeled) observations and an $R_{34}$ MAE of 49 km against the ASCAT observations. The CNN-TC model [44] for estimating intensity is one of the state-of-the-art techniques, with an MAE of approximately 8.5 kt based on 2015−2016 global TCs, while for estimating TC size, the OBTK techniques [59, 60] used in the JTWC have $R_{34}$ MAEs ranging from 48−72 km based on 2014−2016 western North Pacific TCs. Although a homogeneous comparison is not accessible, the DSAT model can predict $V_{max}$ and $R_{34}$ with performances comparable to the state-of-the-art techniques while providing the radial wind profile and more concrete concepts of TC structure that other models cannot provide. *Methods: additional verification* provides further statistical evaluation and profile examples.

**Fig. 2 A case study of DSAT performance for Typhoon Jondati (201815W), including the $V_{max}$ evolution, $R_{34}$ evolution, and wind profile estimates. The track is shown in the upper-right inset.**

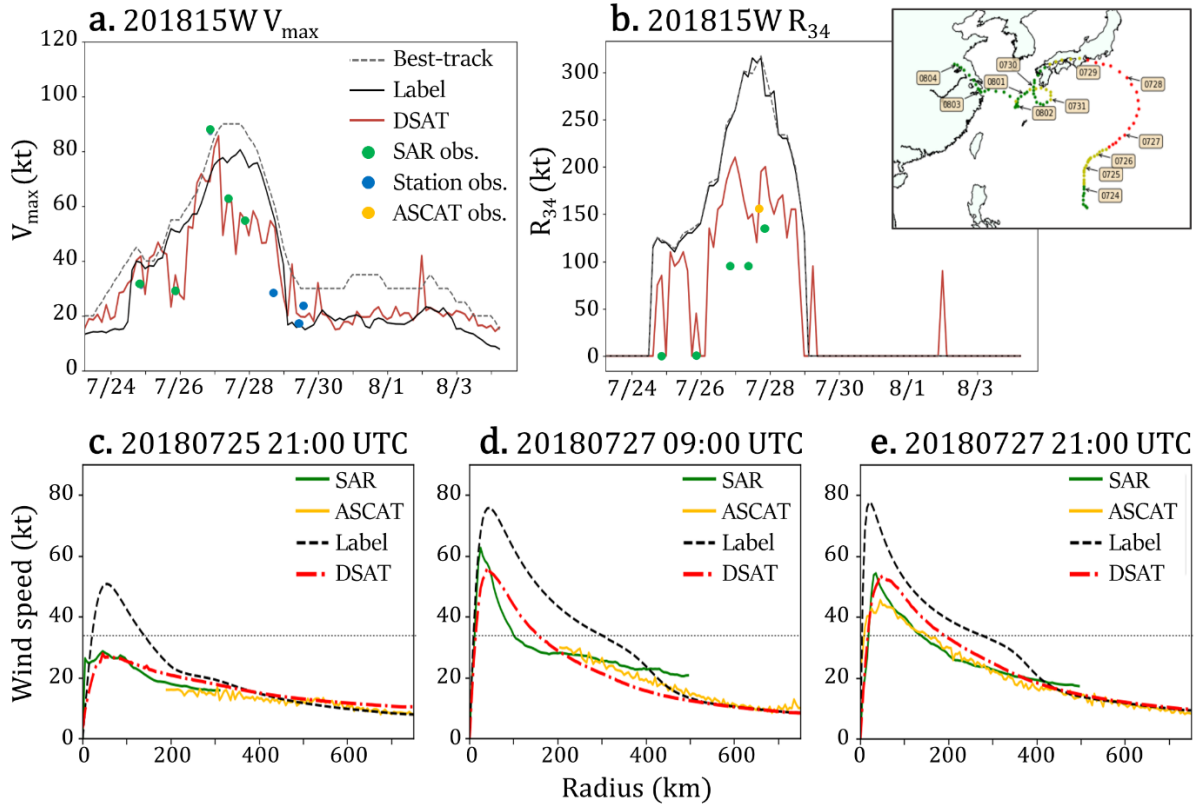

**a** $V_{max}$ evolution based on the DSAT model (red line) and those estimated by the best track (gray dashed line), our labeled data (black line), SAR/ASCAT winds (green and gold dots), and ground stations (blue dots). **b** Corresponding plot for the $R_{34}$ evolution. **c–e** TC wind profiles generated by the DSAT model (dash-dotted red line) when SAR profiles (green line) are available. The ASCAT profiles within +/− 6 h (gold line) and the corresponding labeled profile (dashed black line) are also plotted.

## TC structure trends

A cluster of DSAT models is used to reconstruct homogenized structural parameters for global TCs during 1981–2020 (*methods: left-k-years-out cross-training*). Note that we have excluded samples that might be affected by mid-latitude systems (TC latitude > 35°N/S) and lands (distance to the coast < 50 km) and too weak to be a tropical depression ($V_{max\_DSAT} < 17$ kt or $V_{max\_BT} < 25$ kt). During this period, global TC sample counts show no trends, while samples in the western Northern Hemisphere increased and samples in the Southern Hemisphere decreased (Fig. 3b).

In addition to intensities, we focus on 0–750-km IKEs representing the overall TC energy/circulation:

$$IKE = \int_{v} \frac{1}{2} \rho U^2 \, dV \qquad (Eq. 4),$$

which is computed from the surface wind speeds ($U$) by integrating the kinetic energy per unit volume over the 0–750-km storm domain volume within a 1-m depth ($V$), assuming an air density ($\rho$) of 1 kg m$^{-3}$. Note in Fig. 3c that V$_{max}$ and IKEs are loosely correlated. TCs with similar $V_{max}$ may have substantially different *IKE*s (e.g., 201619S vs. 201517W), while a moderate-intensity TC can still have extreme IKEs (199015W). The questions are how the DSAT intensity trends compare with those based on other datasets [3, 21, 14] and whether other climate trends exist regarding TC structure and IKEs (*methods: trend analysis*).

Our results show (Figs. 3d–f)that in, significantly increasing trends of $V_{max}$ are validated only in terms of the major TC proportion based on the DSAT dataset (Fig. 3d). Although there is an ~13% increase in the DSAT major TC proportion (Fig. 3d, black) in the past four decades, the corresponding values based on best tracks and K2020 data are ~33% and ~25%, respectively. Furthermore, there were no significant trends in DSAT mean $V_{max}$ for typhoons/hurricanes (Fig. 3e), but there was an ~1% per decade increasing trend for best tracks. The DSAT data find no trends in TC rapid intensification (RI, Fig. 3f) and extreme RI (not shown), conflicting with previous best-track studies [3, 14] and modeling studies [5, 21]. The DSAT result suggests that the intensity trends may be exaggerated in analyses based on best tracks due to the underestimated intensities before 2000.

The DSAT reanalysis is one of the first few datasets capable of directly examining the climate trends of TC IKEs (Figs. 4a–c). We first examined the fractional proportion of extremely large IKE TCs (the 95$^{th}$ percentile of all samples) to samples with IKE > 80$^{th}$ percentile (Fig. 4a); this concept is similar to the proportion of major typhoon/hurricane samples to all typhoon/hurricane ($V_{max}$ > ~$q_{80}$) samples examined in Fig. 3d. Figure 4a shows that the proportion of extremely large IKE TCs has significantly increasing trends (~25% over 40 y; see also Extended Fig. E8d), particularly for Northern Hemispheres TCs. Second, the mean IKEs for large-IKE TCs (IKEs > 79 TJ) also show significantly increasing trends (> one standard deviation of the 40-y variability, Fig. 4b). Intriguingly, the proportion of large-IKE samples to typhoon/hurricane samples shows significantly decreasing trends (Fig. 4c), suggesting that the increasing trends regarding IKEs may be not accompanied by the climate trends of TC intensity (see also Extended Fig. E8h). Furthermore, the global mean IKE for small-IKE TCs (IKEs < 79 TJ) has a decreasing trend (Extended Fig. E8g), suggesting that a bimodel component in the IKE distribution was established in the past four decades.

## Discussion and Conclusion

Deep learning provides opportunities to retrieve past failed observations. The TC radial wind profiles reconstructed by the DSAT model provide a second opportunity to examine the structural development, in addition to intensification, of historical TCs and to depict related climate trends, providing insights needed by forecasters and climate modelers who deal with real-world disaster management and sustainable development goals.

Using the 40-year homogenized DSAT dataset, trends of $V_{max}$ and IKE are examined. Although the major TC proportion has increased by ~13% in the past four decades, the intensity trends could be exaggerated based on TC best tracks, and we may overemphasize the role of climate change on intensity [3, 21, 14]. Moreover, the proportion of extremely high IKE samples has increased by ~25%, with an increasing trend of the mean total IKE of high-IKE TCs. Given that the TC damage potential depends more strongly on an integrated measure of the wind fields rather than the pointwise $V_{max}$ [27], an important future research topic is how the climate change affects the growth of TC energy.

It remains challenging to explain the intensity and IKE trends revealed here. Although the warming of SST is predominant, TC migration [1, 61] is found toward areas of lower SSTs and higher VWSs (Figs. 4d–f). TCs in the western Pacific warm pool move toward Taiwan and Japan [62]; TCs in the southern Indian Ocean move toward La Réunion and Madagascar; and decreased samples in the eastern Pacific high-SST area are found (Fig. 4d). A key question is whether and how climate changes in the ocean and general circulation affect the "local-to-TC" (Lagrangian) environment. A naive conjecture is that the poleward migration of TCs could be nature's self-adjustment offsetting TC overintensification. The explanation of IKE trends may be overlaid within future works examining the environmental conditions and internal TC structure evolution (e.g., eyewall replacement cycle [63, 64]) based on the DSAT dataset.

We hope this work will serve as a foundation for new TC observational datasets and methods reconstructing past weather analysis, as well as the greater integration of deep learning and geophysical science in dealing with large meteorological datasets, which makes it possible to promote new research topics and accelerate human adaptation to climate change and weather extremes.

**Fig. 3 40-year (1981–2020) global TC wind profiles reconstructed using the DSAT model. TC structural parameters, such as $V_{max}$ and IKE, can be directly calculated with the reconstructed wind profiles, and their climate trends are examined.**

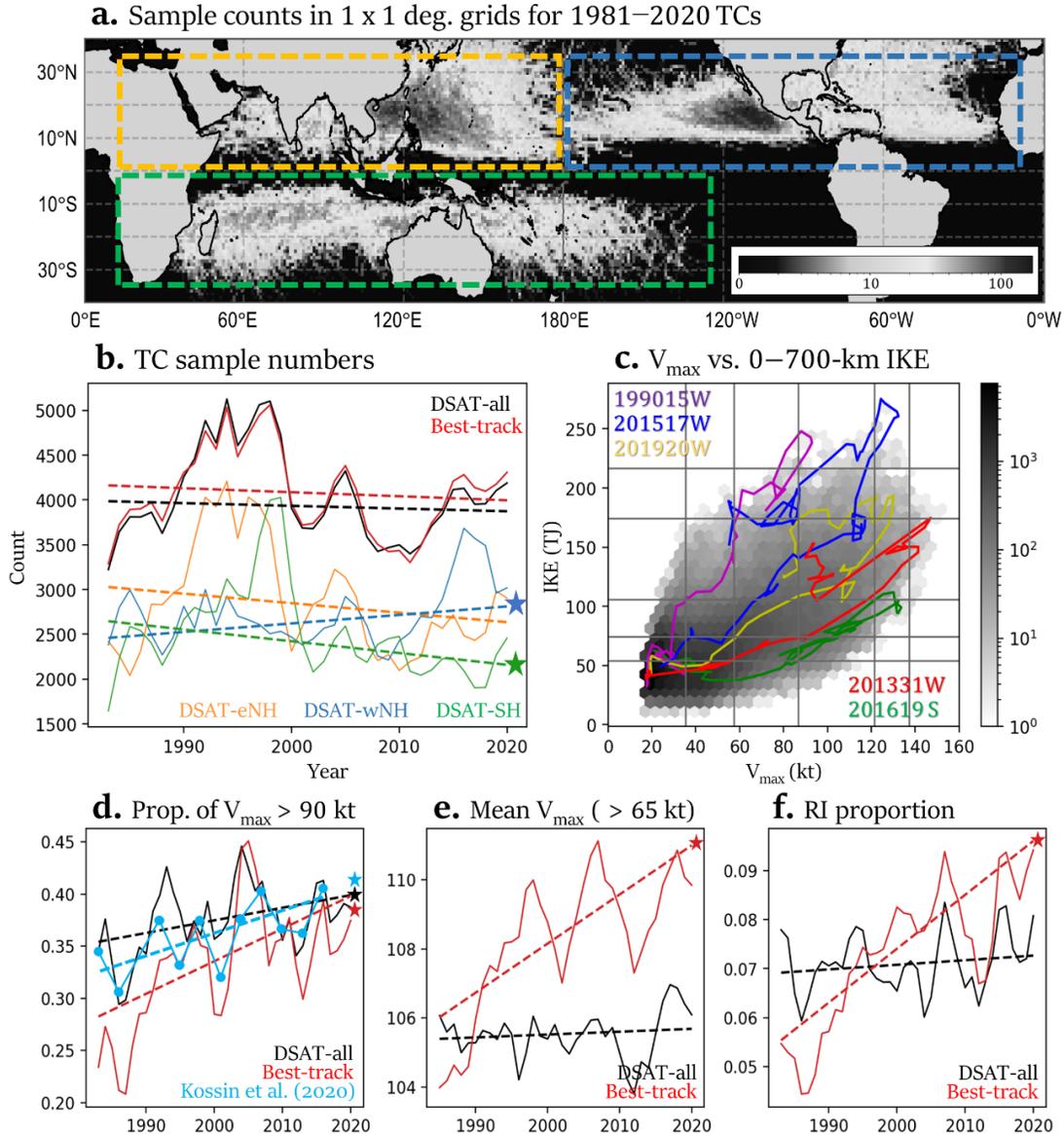

**a** Distribution of 3-hourly sample counts (color bar) in 1 x 1-degree grids for global TCs from 1981 to 2020. **b** Three-yearly averaged annual TC sample numbers from 1981–1983 to 2017–2020 for the global best-track dataset ($V_{max\_BT} > 25$ kt), global DSAT reanalysis dataset ($V_{max\_DSAT} > 17$ kt), eastern NH DSAT data, western NH DSAT data, and SH DSAT data. The corresponding Theil-Sen regression lines (dashed lines) are shown, and the star sign indicates that the trend has passed the Mann–Kendall significance test at the 90% confidence level. **c** Density plot of TC $V_{max\_DSAT}$ and $IKE_{DSAT}$ for all samples; selected TC cases are indicated with colored lines. **d** Same as (b) except for the three-yearly fractional proportion of major typhoon/hurricane samples to all typhoon/hurricane samples for best-track data (red), DSAT data (black), and the K2020 result (aqua, from 1981–1983 to 2015–2017). **e, f** Figures corresponding to (d) for annual mean $V_{max}$ (kt) for typhoon/hurricane samples ($V_{max} > 65$ kt) and fractional proportion of rapid intensification events ($\Delta V_{max\_24h} > 30$ kt).

**Fig. 4 Exploration of the climate trends regarding TC IKEs and discussions regarding the contribution of TC track migration to TC structural trends.**

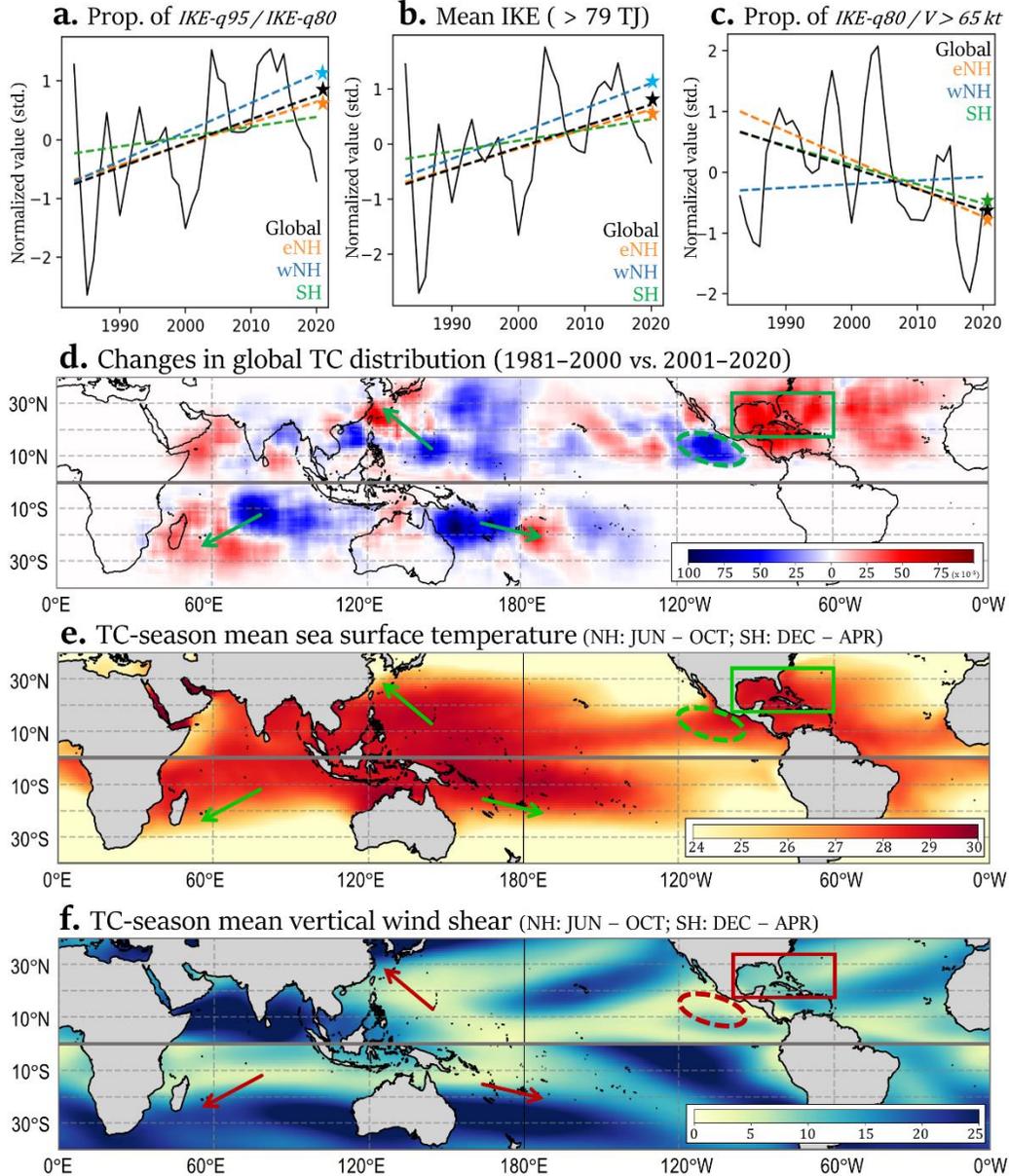

**a–c** Global trends (solid black line) from 1981–1983 to 2017–2020 for (a) the three-year-averaged fractional proportion of TC samples with IKE > 95th percentile of all samples to samples with IKE > 80th percentile, (b) annual mean IKEs of TC samples with IKE > 79 TJ (80th percentile), and (c) proportion of samples with IKE > 80th percentile to typhoon/hurricane samples. Note that Z score normalization was applied to the values. The black, orange, aqua, and green dashed lines are Theil-Sen regression lines for global, eastern NH, western NH, and SH samples, respectively. The star sign indicates that the trend has passed the Mann–Kendall significance test at the 90% confidence level. **d** Differences in global TC sample distribution between years before 2000 and after 2000. **e–f** TC season mean SST (e) and TC season mean deep-layer 850–200 hPa VWS (f) during 1981–2020. The TC season here is referred to as June to October for the NH and December to April for the SH.

## Methods

### Input satellite imagery

Satellite observations have been the primary sources for analyzing TCs over the open ocean [65, 66]. The DSAT model uses satellite images of four channels as the input: IR, WV, VIS, and PMW. A total of 76223 3-hourly time frames were collected during 2004–2018 according to linear-interpolated best-track TC positions. Data within a 1500 x 1500 km$^2$ area of the TC center are collected.

The IR, WV, and VIS are from the Gridded Satellite dataset [67], a long-term dataset of 3-hourly geostationary satellite observations with 0.078 latitude/longitude resolution. The resolution and quality from different satellites have been calibrated, making them suitable data for studying global TC climatology. However, VIS images are only available during daylight hours, so the DSAT model has a special procedure to incorporate temporally heterogeneous data.

The PMW rain rates are derived from the Climate Prediction Centermorphing technique [68], which utilizes low-Earth-orbit microwave satellite observations and combines them with spatial propagation information obtained from geostationary IR data to obtain 3-hourly global coverages of 0.25° latitude/longitude resolution. Previous studies [44, 55] suggested that including PMW images helps improve the accuracy of deep-learning intensity estimation because PMW resolves deep convection in the TC inner core.

### Labeled data

*Source data for labeling.* Best-track data from the Joint Typhoon Warning Center and National Hurricane Center provide the $V_{max\_BT}$, $R_{34\_BT}$, and $RMW_{BT}$ for the DSAT labeling procedure. $R_{34\_BT}$ is the mean radius of the 34-knot wind in the four quadrants. As these data are available only at the 6-hourly synoptic times, linear-interpolated variables are used at 0300, 0900, 1500, and 2100 UTC to match satellite observations. Notably, most of the best-track estimates are based on subjective analysis by forecaster (e.g., Dvorak technique [65]) or satellite consensus [69], rather than in-situ reconnaissance or research observations; thus, best tracks are not homogenized among basins and time and contain considerable uncertainty. Nevertheless, best tracks are currently the most suitable source data for DSAT labeling due to the abundant sample size needed for deep learning.

*The parametric wind model.* Morris and Ruf's parametric wind model [51] is used to fit the TC radial wind profile *V(r)*:

$$V(r) = \frac{2r*(RMW*V_{\max}+\frac{1}{2}f*RMW^2)}{(RMW^2+ar^b)} - \frac{fr}{2},$$

where r is the radius; f is the Coriolis parameter; and a and b are parameters that can be calculated by iteration with $V_{max}$ at *RMW* and the wind speed of 34-kt at $R_{34}$. This wind model is developed based on Cyclone Global Navigation Satellite System (CYGNSS) data that provide observations of surface wind speed under all precipitation conditions. Our previous study [58] also used this wind model for labeling TC profiles and found it to be efficient for reconstructing winds within $R_{34}$ but the deficiency of overestimating the TC outer winds beyond $R_{34}$. Therefore, the current labeling procedure for the DSAT additionally uses the ERA5 winds in the TC outer region to construct the final labeled data for model training.

## Data separation

As shown in Extended Fig. E1, 26917 profiles with $V_{max}$ < 35 kt and 24939 with $V_{max}$ > 35 kt are collected. The training dataset of 36770 samples from 2004–2014 is used to fit the DSAT model weights. The validation dataset of 7535 samples from 2015–2016 was used to find the best hyperparameters that were selected a priori rather than learned during training. We used 7551 testing samples from 2017–2018 to assess the model performance independently for the best-performing model on the validation dataset.

## DSAT model

Deep learning is a subcategory of machine learning and has flourished due to the maturity of GPU computing; it has been applied to research in atmospheric sciences [70, 71, 72], including the identification of severe weather from satellite imagery, statistical downscaling, radar quantitative rainfall forecasting [73, 74], integrating AI modules into numerical weather prediction systems [75], and even developing full-AI-based weather prediction models [76, 77]. The DSAT model (Extended Tab. E1) integrates two special designs: (i) the "polar-convolution filter" [58] suitable for a rotating fluid system and (ii) the GAN-CNN framework [55] handling temporally heterogeneous observations.

*Polar-CNN regressor.* According to the rotational nature of TCs, extracting structural information in polar coordinates is more feasible and physically reasonable than extracting structural information in Cartesian coordinates. The DSAT CNN regressor comprises six polar convolution layers and three dense layers. The polar convolution layers extract useful features from the input images and pass them to the dense layers for regression. A polar-convolution layer is similar to a conventional convolution layer, except for operating in polar coordinates. We project TC images, originally 128x128 in Cartesian coordinates, to 180x103 in polar coordinates, where the first dimension represents the directional angle (2 degrees/point), while the second dimension represents the radius with a 5-km resolution. Note that in polar coordinates, the meaning of a convolution kernel is a sector, instead of a square, with its vertex pointing to the TC center. Detailed hyperparameters are listed in Extended Tab. E1.

After the DSAT regressor outputs a 1 x 151 array of the wind profile, which is used to calculate TC structure parameters (e.g., $V_{max}$, $RMW$, $R_{34}$, and $IKE$), the loss function is calculated to optimize the model weights:

$$L = l_p + \alpha * l_{V_{max}} + \beta * l_{RMW},$$

where $l_p$, $l_{V_{max}}$, and $l_{RMW}$ are the mean absolute errors (MAEs) of the labeled profile, $V_{max}$, and RMW, respectively, and α and β are factors set to 0.0075 and 0.00025 based on a series of experiments. Therefore, $l_p$, $l_{V_{max}}$, and $l_{RMW}$ are optimized simultaneously.

*Rotational argumentation.* When training the DSAT model, images are randomly rotated with respect to the TC center every time before being input to the CNN regressor and the GAN module for data argumentation. This technique [78] helps prevent model overfitting and leads to the decision to omit pooling layers and dropout layers in the model architecture.

*GAN module.* Our previous study [55] showed that a GAN could convert IR and WV images to fake PMW* or fake VIS* images at all times when IR and WV are available, and these PMW*/VIS* can improve the accuracy of $V_{max}$ estimates for the deep learning model. Thus, the DSAT model adopts the GAN module to use temporal-heterogeneous satellite data.

As shown in Fig. 1c and extended Tab. E1, the VIS* GAN generator takes IR, WV, and a scalar parameter *m2n* as the input and uses an adjusted U-Net similar to Pix-2-Pix GAN [79] to generate mimic VIS* images. The generator contains 13 polar convolution/deconvolution layers with skip connections to preserve small-scale image textures. The *m2n* is a 1x1 value indicating the duration (minutes) to noon concerning the VIS observation time; it helps the generator generate VIS* with proper brightness due to sunlight. The trick of *m2n* is that when using the well-trained GAN module, one can set *m2n* to zero and force the generator to generate VIS* images with the brightness at noon. Therefore, the DSAT regressor uses images with consistent brightness for the regression task.

The design of the discriminator follows the idea of PatchGAN [79], which is suitable for satellite images that are continuous, without the concept of objects, foreground, background, and boundaries. The discriminator aims to distinguish real input VIS or fake VIS* and to force the generator to create a better fake. With the generator and discriminator adversarial to each other and trained together, both models can be optimized.

The loss functions of the VIS* GAN include (i) L2 loss of VIS* and VIS, (ii) the generator loss similar to that in patchGAN, and (iii) the CNN regressor loss, which helps the generator create useful features for wind profile regression.

In this study, we also trained a PMW* GAN with input IR/WV images paired with the target PMW images. Extended Fig. E2 showcases the capability of the GAN modules to generate reasonable VIS* and PMW* images. However, as discussed in the next subsection, the final DSAT model only retains the VIS* GAN module (Fig. 1c).

*Optimization and loss analyses*

Four input combinations (Extended Fig E3) are tested: (i) using only IR (IR_only); (ii) using IR and original PMW (IR+PMW); (iii) IR+VIS*, and (iv) IR+PMW*. Note that the IR_only and IR+PMW models only use the CNN regressor to generate the wind profiles, while a three-stage training procedure [55] is adopted for optimizing the hybrid GAN-CNN models using IR+VIS* and IR+PMW* (Extended Fig. E3b):

Stage I: Recalling that the CNN regressor loss is needed for optimizing the GAN module, we pretrained the DSAT regressor with IR and good-quality VIS (or PMW, for the IR+PMW* model) images.

Stage II: The VIS* (PMW*) generator and the discriminator were optimized simultaneously with a typical procedure for optimizing GANs.

Stage III fine-tuned the CNN regressor with the IR and generated VIS* (or PMW*) images.

Briefly, Stages I and II are for training the GAN module, and the last stage is for the CNN regressor; this three-stage training strategy makes the training process more stable.

Extended Fig. E3a compares the learning curves for models using various input combinations. The IR+PMW model reaches the lowest total loss of <5.5 but has a higher risk of overfitting the data, as revealed by its larger amplitude. While the IR+PMW model is better than the IR_only model, it depends on the availability of PMW observations and is not available for TCs before 2004. Furthermore, the IR+PMW* model cannot outperform the IR_only model, and the PMW* generator seems to fail to retrieve useful information from IR and WV images as an alternative PMW-like input.

On the other hand, the IR+VIS* model reaches stable performance after 110 epochs and has a comparable total loss to the IR_only model (Extended Figs. E3a,c–f). Compared with the IR_only model, IR+VIS* better estimates the wind profile (Extended Fig. E3f, $l_p$~4.9 kt) and has comparable performance for $R_{34}$ (slightly better), $V_{max}$ (slightly worse), and RMW (Extended Figs. E3 c–e). Consequently, the IR+VIS* model is selected as the control DSAT model. However, both the IR_only and IR+VIS* models are used for constructing TC structure data for climate research, as described in later sections.

## Independent verification

*The "true ground truth."* Unlike the data strategy (training-validation-testing separation) commonly used in deep learning, this study collects a fourth dataset of high-quality sea surface winds from ASCAT and SAR for independent verification. The limited swath width and sampling frequency inhibit the use of SAR/ASCAT data as labeled data to train deep learning models. However, SAR and ASCAT data are the best "ground truth" for systematic verification of TC inner-core and outer winds, respectively, as the DSAT-labeled data may contain uncertainty caused by the source data.

This study collected 92 SAR observations of TCs during 2017−2018. SAR, C-band high-resolution synthetic aperture radar [80], is the only spaceborne instrument able to probe, at very high resolution and over all ocean basins, the sea surface under extreme weather conditions. In the case of Hurricane Irma (2017), SAR backscatter signals were used to derive high-resolution surface wind estimates up to 75 m s$^{-1}$, with an ~5 m s$^{-1}$ root-mean-square error relative to the winds observed by airborne measurements. Furthermore, the SAR wind profile uniquely describes the inner-core structure, providing measurements of V*max* and RMW (Extended Fig. E4).

This study collected 2055 ASCAT, Advanced Scatterometer [81, 82], profiles for 2017−2018 TCs, and 510 profiles with valid $R_{34}$ were used to verify the DSAT performance in the TC outer region. As the ASCAT cannot observe winds in all quadrants due to the narrow swath of ~500 km, preprocessing of ASCAT data is conducted to obtain the azimuthally mean radial wind profile in the outer region. As shown in Extended Figs. E4b,d, we first calculate the ERA5 surface environmental winds from which the TC vortex is removed ($\vec{V}_{ERA5\_env}$) [53]. Then, using the ASCAT vectors from which $\vec{V}_{ERA5\_env}$ are removed, the azimuthally mean radial wind profile ($\overline{\vec{V}_{scat} - \vec{V}_{ERA5\_env}}$) is obtained. The preprocessed profile reduces the error of the partially scanned ASCAT data for representing the azimuthally mean outer profiles.

*Additional verifications.* Inner-core (<150 km) SAR profiles and outer-region (≥150 km) ASCAT profiles are used to verify the DSAT estimates and the labeled data (Extended Fig E5a). For DSAT estimates, the percentage errors in the inner core are generally less than 50% but show positive biases of 10−25%. As the DSAT model tends not to overestimate $V_{max}$, the errors mainly result from dislocated $RMWs$ and a broader wind profile shape. Moreover, the error distributions of "*DSAT V – SAR V*" have smaller biases than those of "*Label V – SAR V*," implying that the DSAT model can provide even better profiles than the labeled data. Furthermore, the percentage errors are generally less than 25% for outer region winds.

Extended Figs. E5b–e show error distributions and MAEs for subsets with various labeled $V_{max}$ and $R_{34}$ ranges. The DSAT model has larger MAEs in estimating $V_{max}$ (*DSAT $V_{max}$ – SAR $V_{max}$*) for stronger TCs and has larger MAEs for samples with labeled R34 of 200–300 km, which comprise a large portion of strong $V_{max}$ samples. Furthermore, the DSAT model has negative biases in estimating *R34* (*DSAT $R_{34}$ – ASCAT $R_{34}$*) for large TCs, presumably due to the unbalanced data. Finally, although the $V_{max}$ MAE for strong TCs and $R_{34}$ MAE for large TCs are relatively large, the mean absolute percentage errors for estimating $V_{max}$ and $R_{34}$ are approximately 10% and 25%, respectively, implying an acceptable performance of the DSAT estimates.

## Constructing the TC structure reanalysis dataset

*Left-k-years-out cross-training.* When using deep learning models in the prediction phase, data from the training set should not be reinput to yield the model output because the deep neural network is more or less overfit to the training data while achieving the best performance on the validation data (and hopefully the testing data) as it could. As a result, we propose a left-k-years-out cross-training strategy for reconstructing homogenized structure reanalysis for all global TCs during 1981–2020. A cluster of thirteen DSAT models trained/validated with various data separations was used (Extended Fig. E7a). Each model uses 9 years of data from 2004−2016 for training and 4 years of data for validation. For example, models M1, M5, M9, and M12 use data from 2004 as validation data, and thus, we can use the four models to generate reanalysis data for 2004 TCs (Extended Fig. E7a). Thus, combining the thirteen models' outputs based on the validation data can contribute to TC reanalysis during 2004−2016, whereas all models can be used for analyzing TCs before 2004 and after 2016. Finally, the DSAT reanalysis uses the averaged profiles of four models for 2004−2016 TCs and the averaged profiles of all models for TCs in other years. If we randomly select four models to construct wind profiles for TCs not during 2004−2016, consistent results regarding TC climate trends can still be found (not shown).

## Examining climate trends

*Dataset for climate trend analysis.* This study analyzed two TC structure reanalysis datasets(Extended Fig. E7a), including one constructed by IR+VIS* hybrid GAN-CNN models and the other constructed by IR_only DSAT CNN regressors (see Extended Fig. E3 and related contexts). Note that both IR and WV images are needed for constructing the IR+VIS* reanalysis; however, WV images are not available for most of the global TCs before 1995, so the IR+VIS* dataset covers only 1996 to 2020 (Extended Figs. E7b–e, green lines). Comparing the green lines and red lines in Extended Figs. E7b–e, it is suggested that for investigation of TC climate trends mainly based on annually averaged values, the IR_only dataset may lead to consistent results and has the advantage of extending the examination back to the 1980s.

Note in Extended Fig. E7b that the mean best-track $V_{max\_BT}$ is larger than the mean DSAT $V_{max\_DSAT}$ by approximately 17%, highlighting the difference in estimating local maximum sustained wind speeds versus azimuthal mean intensities. Nevertheless, the annual mean DSAT $V_{max}$ estimates are comparable to the training labels (Extended Fig. E7b, dashed black lines). Moreover, comparable evolutions/variabilities in the annual mean *RMW* are found between the labeled and corresponding DSAT estimates (Extended Fig. E7bc). In contrast, the annual mean $R_{34}$ for the label and the corresponding DSAT estimates are not generally identical, presumably due to the higher uncertainty of the best-track $R_{34\_BT}$, but are still considered suitable for climate trend analyses.

In summary, it is justified to use the IR_only dataset to examine the climate trends regarding TC structure, and sixty-two percent of 217,685 records during 1981–2020 are examined after excluding samples that might be affected by mid-latitude systems and lands.

*Trend analysis and statistical significance.* We follow the method in K2020 [4] to conduct trend analysis (Figs. 3c–f and 4a–c). The analyses presented here are based on all DSAT estimates, while some studies [83, 84] only analyzed one representative sample for an individual TC (e.g., lifetime maximum $V_{max}$). We consider all estimates because a TC poses a threat at any time during its lifetime, and a TC with a prolonged development lifespan contributes a higher exceedance probability of strong or large TCs. Furthermore, the trend analysis is based on a 3-year running mean time series, as in K2020. The nonparametric Mann−Kendall test is used to examine the significance of the trends. At the same time, Theil−Sen regression lines represent the tendencies (Extended Tab. E2), which are less sensitive to outliers and provide a robust nonparametric alternative to ordinary least-squares regression. Nevertheless, consistent results are yielded under ordinary least-squares regression.

*Basin variability.* Extended Fig. E8 and Extended Tab. E2 investigate basin variabilities in the TC structure trends. Three regions are examined, as these regions have comparable sample numbers (Figs. 3a,b), including the western Northern Hemisphere (western North Pacific and northern Indian Ocean), eastern Northern Hemisphere (eastern North Pacific and North Atlantic), and Southern Hemisphere (South Pacific and southern Indian Ocean).

Although the current study focuses first on global TC structure trends, it is worth noting that the global trends (Extended Figs. E8a, d–g, black lines) are usually attributed to two regional trends with statistical significance. As significant TC track migrations are found in every basin (Fig. 4d), future studies should examine how the TC structure responds to the local environmental conditions embedded in the regional circulation modified by the changing climate.

## Data Availability

The TC satellite images and labeled data used for developing the DSAT model are temporally available at Google Drive: *https://drive.google.com/drive/folders/1Uu7NYSRA6zCSeaDcZfIraqsXvUhAnX3p* for the review process.

The DSAT structure analysis dataset for 1981−2020 global TCs is also temporally available at Google Drive: *https://drive.google.com/drive/folders/15KniP-NuYdhF9QN7MM1P5BWt36ihizNS*. These data and codes will be moved to a permanent repository following acceptance of the manuscript.

The original geostationary satellite dataset GridSAT is available at

*https://www.ncei.noaa.gov/products/gridded-geostationary-brightness-temperature*.

The original PMW dataset is available at

*https://www.ncei.noaa.gov/products/climate-data-records/precipitation-cmorph*.

TC best-track datasets are available at *https://www.metoc.navy.mil/jtwc/jtwc.html?best-tracks* (JTWC) and *https://www.nhc.noaa.gov/data/#hurdat* (NHC).

The SAR and ASCAT datasets are available at

*https://www.star.nesdis.noaa.gov/socd/mecb/sar/sarwinds_tropical.php* and

*https://www.remss.com/missions/ascat/*.

## Code Availability

Samples of the compiled dataset for model training and validation and the source code of the DSAT model are available to the public and can be downloaded at *https://github.com/BoyoChen/Climate-Trends-of-TC-Revealed-by-DL*.

## Acknowledgments

The authors appreciate the feedback from forecasters of the Central Weather Bureau, Taiwan. Computational resources for this study were mainly provided by the Center for Weather Climate and Disaster Research, National Taiwan University. This project was funded by Grant MOST 109-2625-M-002-021 and MOST 110-2625-M-002-021 of the Ministry of Science and Technology, Taiwan, and Project 1102056E of the Central Weather Bureau, Taiwan.## Ethics declarations

**Competing interests.** The authors declare that they have no competing interests.

## References

[1] Kossin, J. P., Emanuel, K. A., & Vecchi, G. A. (2014). The poleward migration of the location of tropical cyclone maximum intensity. Nature, 509(7500), 349-352.

[2] Song, J., & Klotzbach, P. J. (2018). What has controlled the poleward migration of annual averaged location of tropical cyclone lifetime maximum intensity over the western North Pacific since 1961?. Geophysical Research Letters, 45(2), 1148-1156.

[3] Knutson, T., Camargo, S. J., Chan, J. C., Emanuel, K., Ho, C. H., Kossin, J., ... & Wu, L. (2019). Tropical cyclones and climate change assessment: Part I: Detection and attribution. Bulletin of the American Meteorological Society, 100(10), 1987-2007.

[4] Kossin, J. P., Knapp, K. R., Olander, T. L., & Velden, C. S. (2020). Global increase in major tropical cyclone exceedance probability over the past four decades. Proceedings of the National Academy of Sciences, 117(22), 11975-11980.

[5] Knutson, T., Camargo, S. J., Chan, J. C., ... & Wu, L. (2020). Tropical cyclones and climate change assessment: Part II: Projected response to anthropogenic warming. Bulletin of the American Meteorological Society, 101(3), E303-E322.

[6] Schreck III, C. J., Knapp, K. R., & Kossin, J. P. (2014). The impact of best track discrepancies on global tropical cyclone climatologies using IBTrACS. Monthly Weather Review, 142(10), 3881-3899.

[7] Emanuel, K., Caroff, P., Delgado, S., Guard, C. C., Guishard, M., Hennon, C., ... & Vigh, J. (2018). On the desirability and feasibility of a global reanalysis of tropical cyclones. Bulletin of the American Meteorological Society, 99(2), 427-429..

[8] Masson-Delmotte, V., Zhai, P., Pörtner, H. O., Roberts, D., Skea, J., & Shukla, P. R., Global Warming of 1.5 C: IPCC special report on impacts of global warming of 1.5 C above pre-industrial levels in context of strengthening response to climate change, sustainable development, and efforts to eradicate poverty., Cambridge University Press, 2022.
25

# Extended Data Figures and Tables

**Extended Data Figure E1: Flow chart for dataset preparation. Fifteen years of data were used to construct the training, validation, and testing datasets for the DSAT model.**

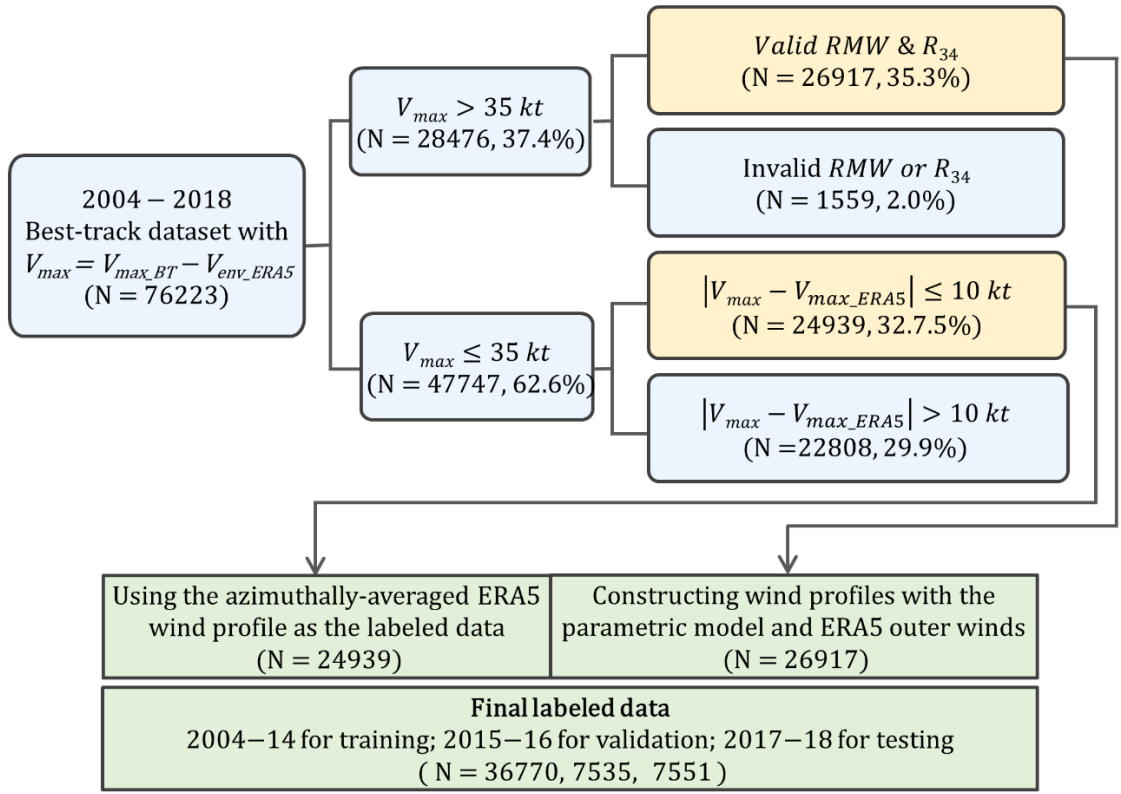

For best-track records with valid $V_{max}$, $R_{34}$, and RMW, labeled wind profiles are created by connecting the parametric model winds and ERA5 outer winds, while 24939 tropical depression wind profiles are directly labeled with the azimuthally averaged ERA5 surface winds with respect to the best-track center.

**Extended Data Table E1: Model designs and hyperparameters for the DSAT GAN module (generator and discriminator) and the DSAT polar-CNN regressor.**

| | | | Generator structure | | | | | |
|---|---|---|---|---|---|---|---|---|
| Layer | Oper. | Input dm. | Kernel dim. | Stride | Filter num. | Batch normal. | Activ. | Skip connection |
| L1 | Conv. | (180,103,2) | (4,3) | (2,2) | 32 | N | Leaky | To L13 |
| L2 | Conv. | (90,52,32) | (4,3) | (2,2) | 64 | Y | Leaky | To L12 |
| L3 | Conv. | (45,26,64) | (4,3) | (2,2) | 128 | Y | Leaky | To L11 |
| L4 | Conv. | (23,13,128) | (4,3) | (2,2) | 256 | Y | Leaky | To L10 |
| L5 | Conv. | (12,7,256) | (4,3) | (2,2) | 256 | Y | Leaky | To L9 |
| L6 | Conv. | (6,4,256) | (4,3) | (2,2) | 256 | Y | Leaky | - |
| L7 | | Concatenating 11 constant 3x2 arrays of auxiliary feature maps: {basin*6, lon., lat., m2n., day-of-year*2} | | | | | | |
| L8 | Deconv. | (3,2,267) | (4,3) | (2,2) | 256 | Y | Relu | - |
| L9 | Deconv. | (6,4,512) | (4,3) | (2,2) | 512 | Y | Relu | - |
| L10 | Deconv. | (12,7,512) | (4,3) | (2,2) | 384 | Y | Relu | - |
| L11 | Deconv. | (23,13,384) | (4,3) | (2,2) | 192 | Y | Relu | - |
| L12 | Deconv. | (45,26,192) | (4,3) | (2,2) | 96 | Y | Relu | - |
| L13 | Deconv. | (90,52,96) | (4,3) | (2,2) | 32 | Y | Relu | - |
| L14 | Conv. | (180,103,32) | (4,3) | (1,1) | 1 | N | Relu | - |
| Output Dim. = (180,103,1) | | | | | | | | |

| | | | Discriminator structure | | | | |
|---|---|---|---|---|---|---|---|
| Layer | Oper. | Input dm. | Kernel dim. | Stride | Filter num. | Batch normal. | Activ. |
| L1 | Conv. | (180,103,1) | (4,3) | (2,2) | 32 | N | Leaky |
| L2 | Conv. | (90,52,32) | (4,3) | (2,2) | 64 | Y | Leaky |
| L3 | Conv. | (45,26,64) | (4,3) | (2,2) | 128 | Y | Leaky |
| L4 | Conv. | (23,13,128) | (4,3) | (2,2) | 256 | Y | Leaky |
| L5 | Conv. | (25,15,256) | (4,3) | (2,2) | 1 | Y | Relu |
| Output Dim. = (22, 12, 1) | | | | | | | |

| | | | Polar-CNN regressor structure | | | | |
|---|---|---|---|---|---|---|---|
| Layer | Oper. | Input dim. | Kernel dim. | Stride | Filter num. | Batch normal. | Activ. |
| L1 | Conv. | (180,103,2) | (4,3) | (2,2) | 16 | Y | Relu |
| L2 | Conv. | (90,52,16) | (4,3) | (2,2) | 32 | Y | Relu |
| L3 | Conv. | (45,26,32) | (4,3) | (2,2) | 64 | Y | Relu |
| L4 | Conv. | (23,13,64) | (4,3) | (2,2) | 128 | Y | Relu |
| L5 | Conv. | (12,7,128) | (4,3) | (2,2) | 256 | Y | Relu |
| L6 | Conv. | (6,4,256) | (4,3) | (2,2) | 512 | Y | Relu |
| | | Flatten to (3,2,512) and concatenate 10 auxiliary variables: {basin*6, local-time*2, day-of-year*2} | | | | | |
| L7 | Dense | 3082 | - | - | 256 | Y | Relu |
| L8 | Dense | 256 | - | - | 64 | Y | Relu |
| L9 | Dense | 64 | - | - | 151 | Y | Relu |
| Output Dim. = (151) | | | | | | | |

The GAN generator has a 13-layer U-Net structure with convolution ("Conv.") and deconvolution ("Deconv.") layers. The GAN discriminator has five convolution layers. The polar-CNN regressor consists of six convolution layers and three dense layers. The table shows the input dimension of each layer, the kernel dimension of the convolution filter, the stride for scanning, the number of filters, the use of batch normalization, and the activation function.

**Extended Data Figure E2: The GAN module can using IR and WV images to generate fake VIS* and PMW* images.**

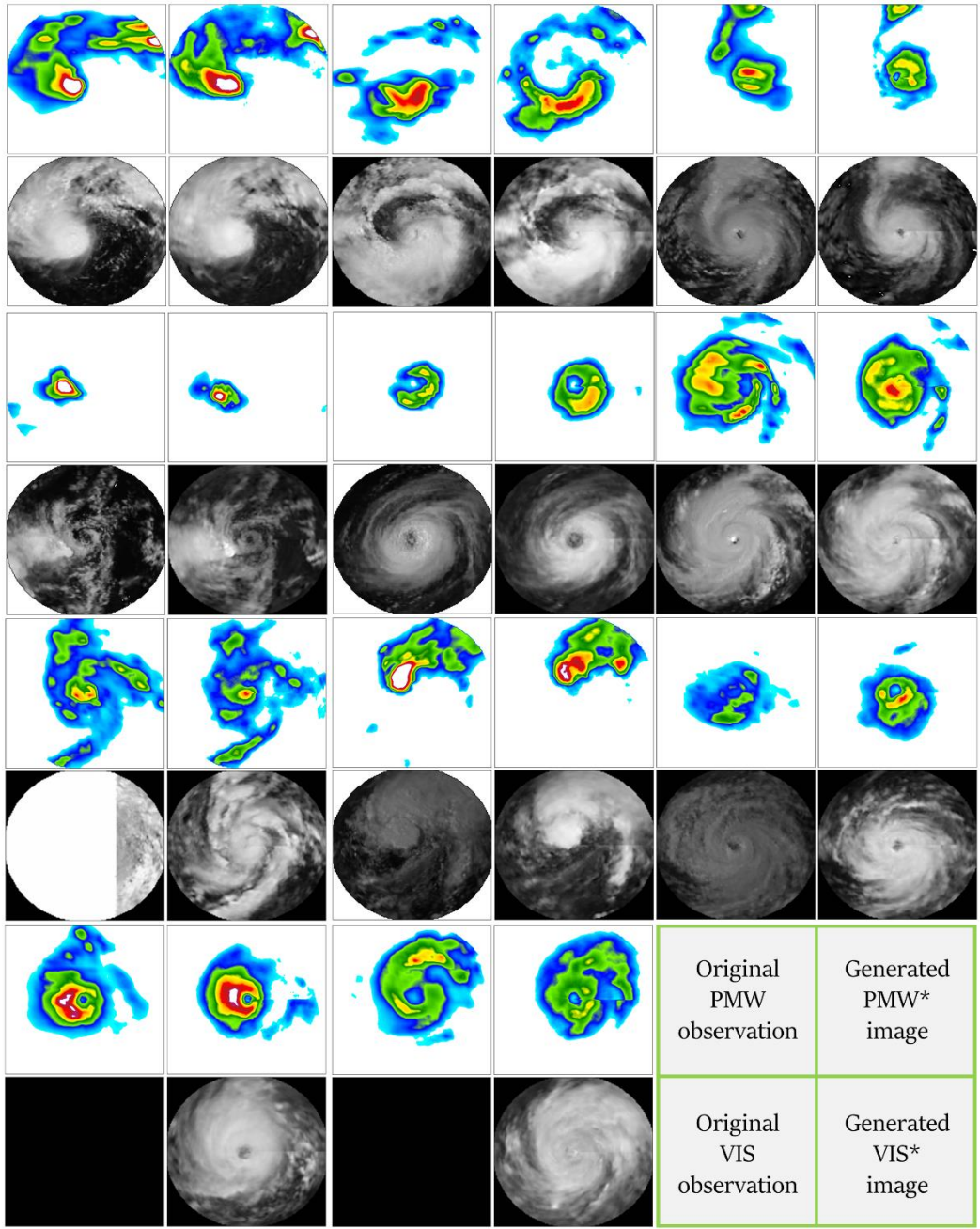

For every 2x2 subpanel, the upper left is the original PMW image, the lower left is the original VIS image, and the right side shows the corresponding generated PMW* and VIS* images. Examples in the third and fourth rows showcase that the GAN module generates VIS* during the evening and nighttime, while examples in the third column showcase that the GAN module generates reasonable PMW* when the original PMW has poor quality, resolving the inner-core structure.

**Extended Data Figure E3: The three-stage training strategy and performance evaluations for models using various input combinations: IR_only, IR+PMW, IR+VIS*, and IR+PMW*.**

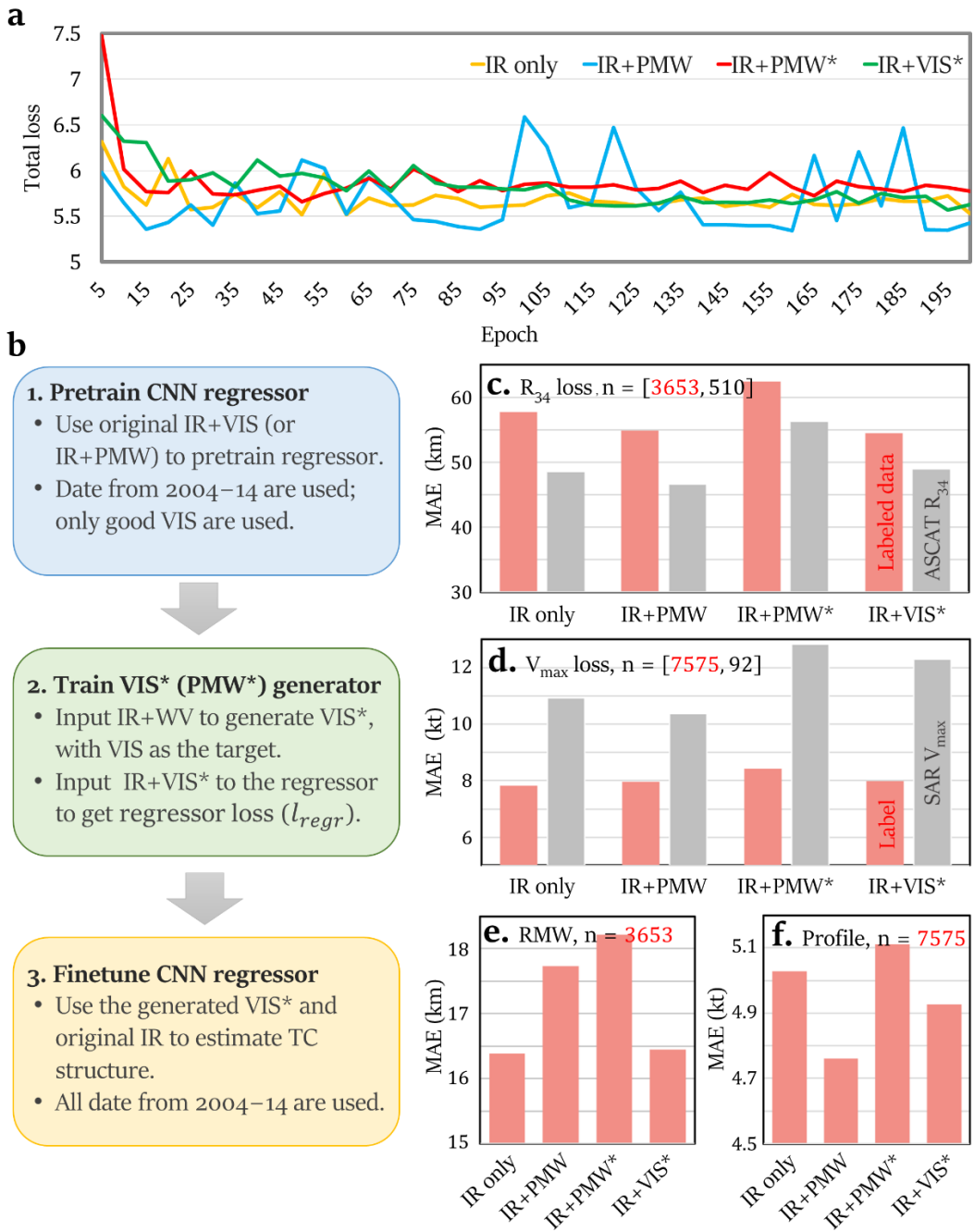

**a** Learning curves for models using various input combinations. The Y-axis is the total loss of the DSAT regressor. **b** The three-stage process for training the hybrid GAN-CNN model. **c**–**f** For the testing dataset, the $R_{34}$ MAE (c), $V_{max}$ MAE (d), RMW MAE (e), and profile MAE (f) for the various models with respect to the labeled data (red bars) and ASCAT $R_{34}$ or SAR $V_{max}$ (gray bars) from the independent verification dataset.

**Extended Data Figure E4: The "true ground truth" in this study: independent ASCAT and SAR satellite winds. A total of 510 ASCAT scans and 92 SAR scans are used to verify the DSAT model.**

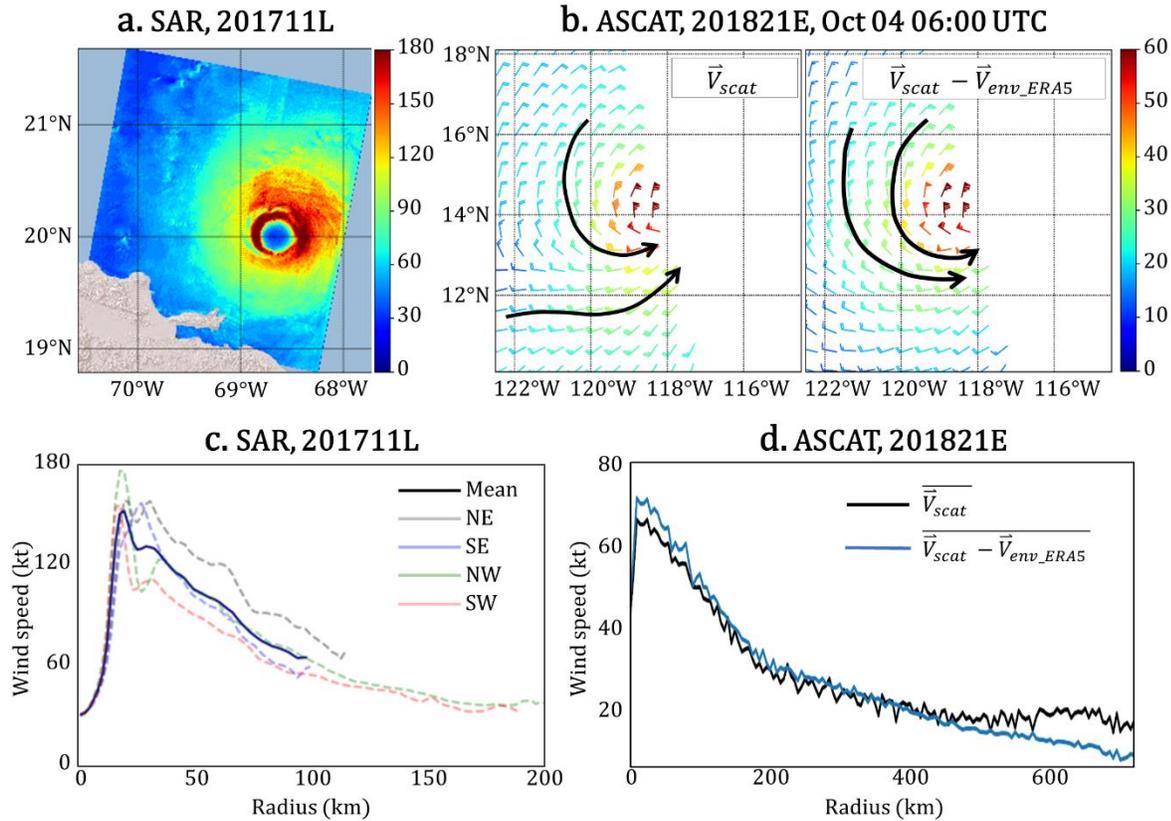

**a, c** An example of SAR observations of Hurricane Irma (201711L) at 10:30 UTC on 07 September 2017 includes the 2-D SAR surface winds (color bar, knot), azimuthal mean wind profile (solid dark blue line) and wind profiles for the four quadrants. The azimuthal mean wind profile is used to verify the DSAT performance in the TC inner core. **b, d** An example of an ASCAT observation of TC 201821E at 06:00 UTC on 04 October 2018. (b) The original ASCAT observation ($\vec{V}_{scat}$, left) and wind vectors from which the ERA5 environmental winds are removed ($\vec{V}_{scat} - \vec{V}_{ERA5\_env}$, right). (d) Azimuthal mean radial wind profiles for $\vec{V}_{scat}$ and $\vec{V}_{scat} - \vec{V}_{ERA5\_env}$ winds. The profile $\overline{\vec{V}_{scat} - \vec{V}_{ERA5\_env}}$ is used to verify the DSAT performance in the outer region.

**Extended Data Figure E5: Statistical verification of the DSAT-estimated wind profiles, $V_{max}$, and $R_{34}$ with respect to the independent SAR (92 samples) and SCAT (510 samples) observations for 2017–2018 TCs in the testing dataset.**

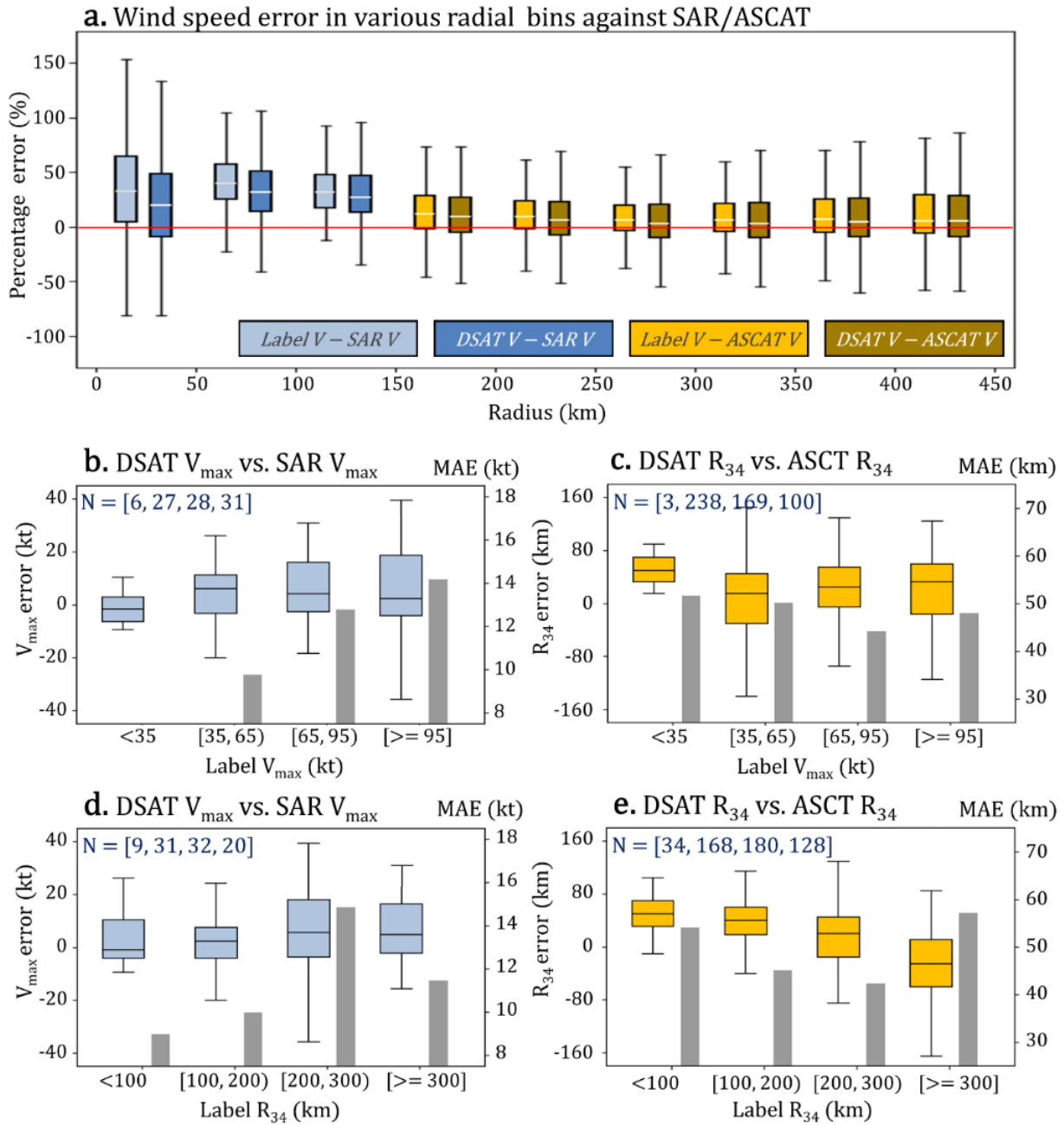

**a** Box-and-whisker plots show the percentage error distributions to the SAR/ASCAT winds for the labeled data (Label V) and DSAT profiles (DSAT V). Note that here, the "truth" consists of the SAR V in the inner core (r < 150 km) and the ASCAT V in the outer region. **b, d** The error distributions (box-and-whisker plot, left Y-axis, kt) and MAEs (bar graph, right Y-axis, kt) between the DSAT $V_{max}$ and the SAR $V_{max}$ (92 samples) for subsets of various ranges of labeled $V_{max}$ (b) and labeled $R_{34}$ (d), respectively. **c, e** Same as b, d, except for verification between DSAT $R_{34}$ and ASCAT $R_{34}$ (510 samples).

**Extended Data Figure E6: This figure shows another eight examples of verifying the DSAT-analyzed TC profiles against the independent SAR and ASCAT profiles.**

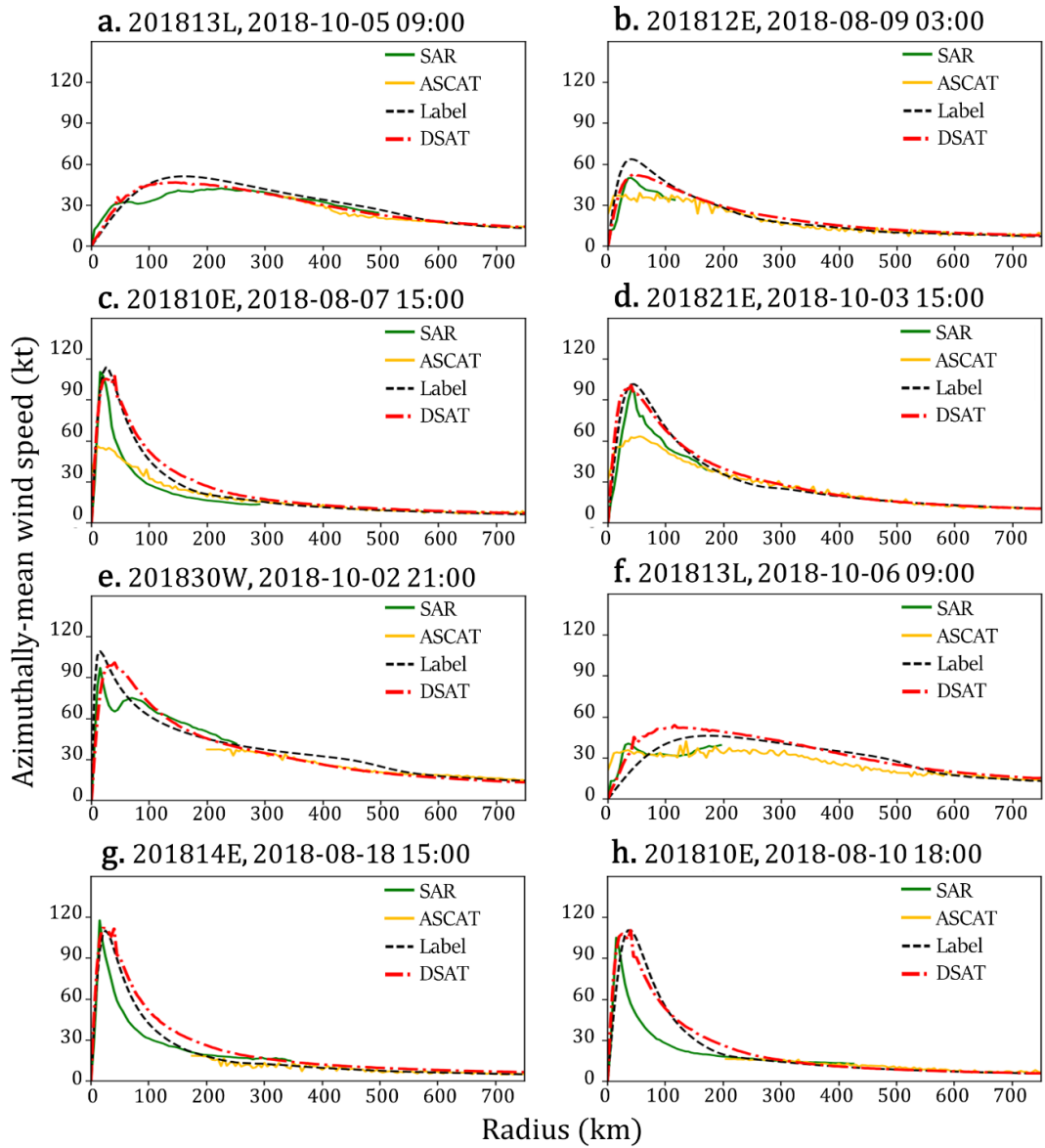

Each plot shows the TC wind profile generated by the DSAT (dash-dotted red line), the SAR profile (green line), the ASCAT profile within +/− 6 h (gold line), and the corresponding label profile (black-dashed line). The TC ID and time (UTC) are also listed. These examples clearly show that the SAR provides high-quality inner-core profiles, and the ASCAT profile is useful for verifying the TC outer winds while tending to underestimate the inner-core winds. Moreover, the DSAT model well captures the wind profiles of these TCs with reasonable $V_{max}$, RMW, and decaying outer winds. However, the cases in (e) and (f) indicate that the model cannot reveal a secondary wind peak, while the cases in (g) and (h) show the deficiency of the DSAT model in generating extremely small amounts of RMW. These limitations are expected because the labeled data can neither exhibit secondary wind peaks nor very small RMWs because of the usage of TC best tracks.

**Extended Data Figure E7:** Left-k-years-out cross-training for using the DSAT model to reconstruct homogenized structure reanalysis for all global TCs during 1981–2020. The comparison between using only IR data and IR+VIS* data justifies the usage of deep-learning-generated data to examine the climatic trends of TC structure.

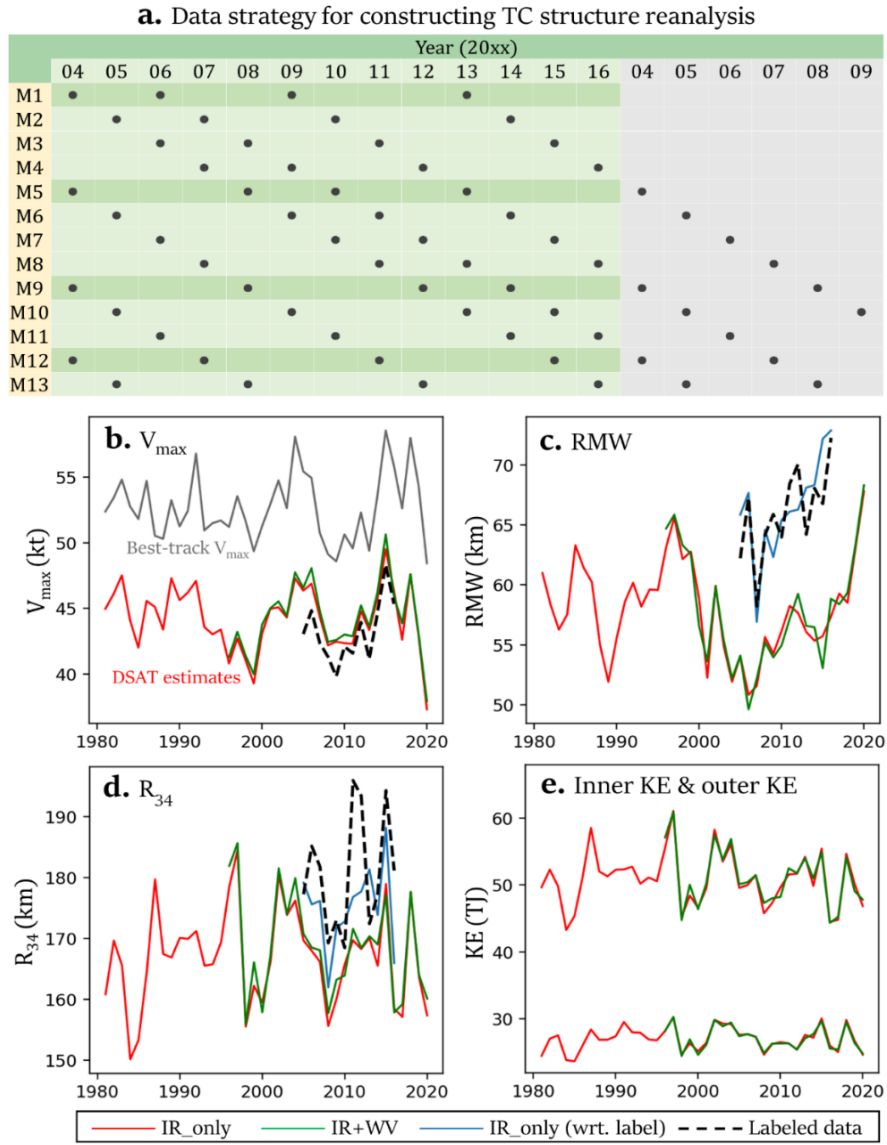

**a** The training-validation data separation for training a cluster of thirteen models (M1–M13) is applied to construct TC structure reanalysis from 1981 to 2020. Data from 2004–2016 are used for training. The dark gray dots indicate that data from a particular year belong to the validation dataset of a particular model. **b–e** The annual averaged $V_{max}$ (b), RMW (c), $R_{34}$ (d), and inner IKE (< 180 km, the mean of all available $R_{34}$) and outer (180–750 km) IKE (e) for DSAT-reanalyzed datasets based on only IR images (red lines, the IR_only models) and IR plus WV images (green lines, the IR+VIS* models), best-track records (gray lines) and the labeled dataset (dashed black lines). Note that only a portion of all samples during 2004–2016 is used to calculate the labeled data, and TCs with $V_{max} < 35$ kt have no $R_{34}$/RMW labels. Therefore, we also plot the annual averages of the IR_only estimates (blue lines) with respect to the labeled data (dashed black lines) for homogeneous comparisons.

**Extended Data Figure E8: Examining basin variabilities of climate trends of TC structure based on the homogenized DSAT reanalysis data.**

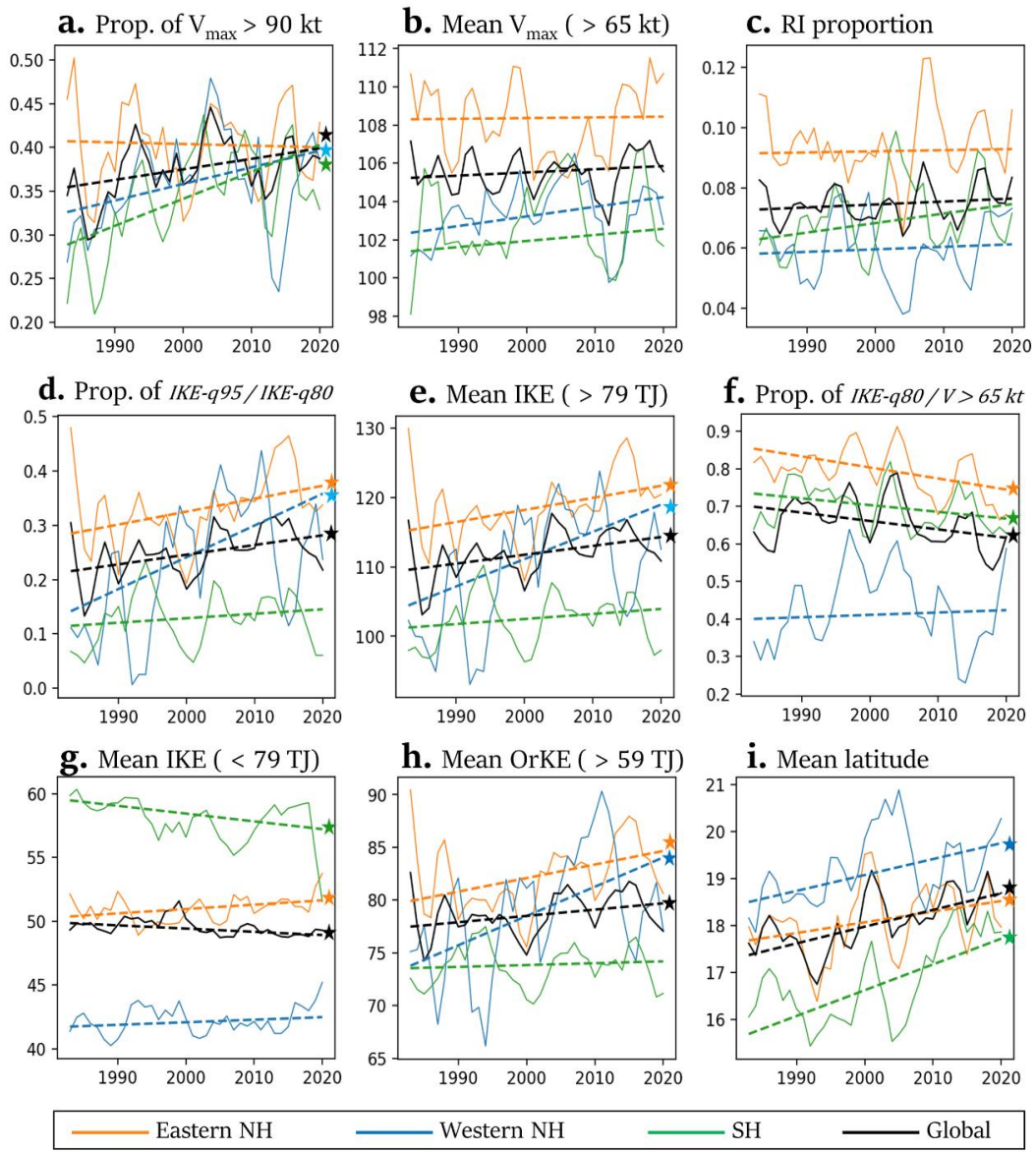

**a** Three-year averaged annual fractional proportion of major typhoon/hurricane samples from 1981–1983 to 2017–2020 for the DSAT reanalysis for global samples (black), eastern NH samples (orange), western NH samples (aqua), and SH samples (green). The corresponding Theil-Sen regression lines (dashed lines) are shown, and the star sign indicates that the trend has passed the Mann–Kendall significance test at the 90% confidence level. **b-i** Similar plots to (a), except for (b) mean $V_{max}$ for typhoon/hurricane samples, (c) fractional proportions of rapid intensification events, (d) proportion of samples with IKE > 95th percentile to samples with IKE > 80th percentile, (e) mean IKEs of TC samples with IKE > 79 TJ (80th percentile), (f) proportion of samples with IKE > 80th percentile to typhoon/hurricane samples, (g) mean IKE for samples with IKE < 79 TJ, (h) mean outer KE for samples with outer KE > 59 TJ, and (i) TC latitudes.

**Extended Data Table E2: Statistics of climate trends of TC structure based on the homogenized DSAT reanalysis and best-track data.**

|  |  | DSAT | | | | Best-track |
|---|---|---|---|---|---|---|
|  |  | Global | eNH | wNH | SH | Global |
| Sample Number | Trend | no trend | no trend | *increasing* | *decreasing* | no trend |
|  | P | 0.763 | 0.227 | **0.083** | **0.050** | 0.669 |
|  | Slope | -3.028 | -10.615 | 9.500 | -13.152 | -4.471 |
| Proportion of $V_{max} > 90$ kt | Trend | *increasing* | no trend | *increasing* | *increasing* | *Increasing* |
|  | P | **0.017** | 0.821 | **0.018** | **<0.001** | **<0.001** |
|  | Slope | 0.0012 | -0.00018 | 0.00192 | 0.00310 | 0.00311 |
| Mean $V_{max}$ (> 65 kt) | Trend | no trend | no trend | *increasing* | no trend | *increasing* |
|  | P | 0.302 | 0.860 | **0.046** | 0.352 | **<0.001** |
|  | Slope | 0.01662 | 0.00371 | 0.05041 | 0.03167 | 0.13972 |
| RI Proportion ($\Delta V > 30$ kt) | Trend | no trend | no trend | no trend | *increasing* | *increasing* |
|  | P | 0.247 | 0.880 | 0.725 | **0.074** | **<0.001** |
|  | Slope | 0.00009 | 0.00003 | 0.00010 | 0.00033 | 0.00110 |
| Proportion of IKE > q95 over IKE > q80 | Trend | *increasing* | *increasing* | *increasing* | no trend | N/A |
|  | P | **0.005** | **0.015** | **<0.001** | 0.308 | N/A |
|  | Slope | 0.00178 | 0.00238 | 0.00585 | 0.00082 | N/A |
| Proportion of IKE > q80 over $V_{max} > 65$ kt | Trend | *decreasing* | *decreasing* | no trend | *decreasing* | N/A |
|  | P | **0.070** | **0.013** | 0.529 | **0.052** | N/A |
|  | Slope | -0.00224 | -0.00295 | 0.00063 | -0.00180 | N/A |
| Mean IKE (> 79 TJ) | Trend | *increasing* | *increasing* | *increasing* | no trend | N/A |
|  | P | **0.006** | **0.004** | **0.001** | 0.290 | N/A |
|  | Slope | 0.12761 | 0.17385 | 0.39311 | 0.07251 | N/A |
| Mean IKE (< 79 TJ) | Trend | *decreasing* | *increasing* | no trend | *decreasing* | N/A |
|  | P | **0.003** | **0.044** | 0.237 | **0.008** | N/A |
|  | Slope | -0.02535 | 0.03415 | 0.02005 | -0.06102 | N/A |
| Mean OrKE (> 59 TJ) | Trend | *increasing* | *increasing* | *increasing* | no trend | N/A |
|  | P | **0.092** | **0.025** | **0.002** | 0.597 | N/A |
|  | Slope | 0.05977 | 0.12833 | 0.27738 | 0.01744 | N/A |
| Mean latitude | Trend | *increasing* | *increasing* | *increasing* | *increasing* | *increasing* |
|  | P | **<0.001** | **0.049** | **0.002** | **<0.001** | **<0.001** |
|  | Slope | 0.03562 | 0.02276 | 0.033891 | 0.05490 | 0.03562 |

Trends, Mann–Kendall P values, and the slopes of Theil-Sen regression lines are shown for corresponding variables to Figs. 3b, d–f, Figs. 4a–c, and Fig. E8 for various regions based on DSAT reanalysis and best-track data. Bold fonts indicate statistical significance.